\documentclass[12pt,preprint]{aastex}
\usepackage{graphicx}
\begin{document} 

\title{The Size Difference Between Red And Blue Globular Clusters Is NOT Due To Projection Effects}
\author{Jeremy J. Webb, William E. Harris, Alison Sills}
\affil{Department of Physics and Astronomy, McMaster University, Hamilton ON L8S 4M1, Canada}
\email{webbjj@mcmaster.ca}

\begin{abstract}

Metal-rich (red) globular clusters in massive galaxies are, on average, smaller than metal-poor (blue) globular clusters. One of the possible explanations for this phenomenon is that the two populations of clusters have different spatial distributions. We test this idea by comparing clusters observed in unusually deep, high signal-to-noise images of M87 with a simulated globular cluster population in which the red and blue clusters have different spatial distributions, matching the observations. We compare the overall distribution of cluster effective radii as well as the relationship between effective radius and galactocentric distance for both the observed and simulated red and blue subpopulations. We find that the different spatial distributions does not produce a significant size difference between the red and blue subpopulations as a whole, or at a given galactocentric distance. These results suggest that the size difference between red and blue globular clusters is likely due to differences during formation or later evolution.

\end{abstract}


\section{Introduction \label{Introduction}}

Globular cluster populations within many types of galaxies have colour distributions which are bimodal \citep[e.g.][] {zepf93, larsen01, harris09b, peng06}. This split in colour is assumed to be due to a wide range in cluster metallicity, with red clusters being metal rich compared to blue clusters \citep[e.g.][] {zepf93, brodie06}. The fact that this bimodality is seen in all but the smallest galaxies indicates that globular clusters may share a common formation mechanism and have similar stages of evolution from one galaxy to the next. 

An interesting difference between these two sub-populations has emerged from a long list of observational studies which find that blue globular clusters have effective radii ($r_h$) that are on average larger than red globular clusters by approximately $20 \%$ ($\sim 0.4$ pc) \citep[e.g.][]{kundu98, kundu99, larsen01, jordan05, harris09b, harris10, paolillo11, blom12, woodley12}.

There exist several competing ideas regarding the source of this size difference. \citet{larsen03} suggested that the size different is not intrinsic, and is due to a projection effect of the different spatial distributions of the red and blue subsystems. Many studies show that the red sub-population tends to be more centrally concentrated in the host galaxy than the blue sub-population \citep[e.g.][] {larsen01, forbes06, harris09, harris09b}. From basic tidal theory it is known that the tidal radius ($r_t$) of a globular cluster is proportional to $M_{cl}^{\frac{1}{3}} R_{gc}$ \citep{vonhoerner57}, where $M_{cl}$ is the cluster's mass and $R_{gc}$ its galactocentric distance. Therefore, since red globular clusters typically orbit in a stronger tidal field than blue clusters, they will be subject to increased tidal stripping and may then have smaller tidal radii. If the structural properties of the clusters, such as their central concentrations \textit{c}, do not vary systematically with $R_{gc}$, then the mean effective (or half-light) radius $r_h$ should also increase with $R_{gc}$. This statement is supported by many observational studies which find a general increase in cluster size with galactocentric distance \citep[e.g.][] {vdbergh96, spitler06, gomez07, harris09b, woodley12}. Therefore, \citet{larsen03} concluded that blue clusters are only larger than red clusters at all projected distances because they have larger three dimensional galactocentric distances. However, \citet{harris09b} showed that for six giant elliptical galaxies the ratio between mean red cluster size and mean blue cluster size did not change with galactocentric distance. This was also observed by \citet{paolillo11} in NGC 1399 and \citet{blom12} in NGC 4365. One would expect that at larger galactocentric distances, projection effects would be negligible and that red and blue clusters would have similar sizes (see \citet{larsen03}). These findings strongly suggest that the size difference between red and blue clusters is intrinsic. Studies which support the projection effect hypothesis include observations of the Sombrero galaxy by \citet{spitler06}, however this is not a giant elliptical galaxy like M87.

Explanations in the literature which propose that the size difference between red and blue clusters is real suggest it is the result of red and blue clusters having different dynamical and stellar evolution histories. \citet{jordan04} suggested that stellar evolution time scales are dependent on metallicity such that many internal relaxation times of dynamical evolution would result in metal-rich clusters being smaller than metal-poor clusters. This explanation is supported by recent N-body simulations by \citet{schulman12} of young clusters and \citet{sippel12} of clusters evolved to a Hubble time. However, a recent paper by \citet{downing12} suggests instead the size difference could be attributed to black holes in blue clusters heating their central regions. Finally, \citet{jordan05} and \citet{harris09b} proposed that metal-rich clusters may have undergone more rapid cooling and contraction while they were gaseous protoclusters, before the majority of their stars had formed. \citet{kundu98} also suggested the discrepancy is due to differences in the formation of metal-poor and metal-rich globular clusters.  

The purpose of this study is to further explore the possibility that in M87 the observed size difference is due to the projection of the different spatial distributions of red and blue globular clusters. While this is simply one possible explanation for the observed size difference, confirming or ruling out this possibility will contribute to determining whether the size difference is intrinsic or artificial. In a previous paper \citep{webb12} (hereafter WSH12), we simulated a globular cluster population orbiting in the tidal field of M87 and reproduced the observed relationship between cluster size and projected galactocentric distance. In our simulation, the only intrinsic difference between red clusters and blue clusters is their assumed spatial distribution. Therefore, we can compare characteristics of the observed and simulated red and blue cluster populations and determine if the projection of the red and blue cluster spatial distributions can explain the observed size difference. More specifically, if the size difference is truly due to the tidal truncation of differing spatial distributions then our simulation will show that the mean red cluster size is smaller than the mean blue cluster size and that red clusters are smaller than blue clusters at all projected galactocentric distance. 

\section{Observations \label{stwo}}

The HST ACS/WFC Archive images we use in this study are from program GO-10543 (PI Baltz). The co-added composite exposures in each filter are described in detail in \citet{bird10}, constructed through use of the APSIS software \citep{blakeslee03}. While these are the same raw images used by \citet{madrid09}, \citet{peng09} and \citet{waters09}, as discussed in WSH12 our images have an improved scale of 0\farcs 025~px$^{-1}$ (half the native pixel size of the camera). This produced an improvement in the effective spatial resolution of the data compared with all previous studies \citep[see][ and WSH12 for a more detailed description]{bird10}.

\begin{figure}[tbp]
\centering
\includegraphics[width=\columnwidth]{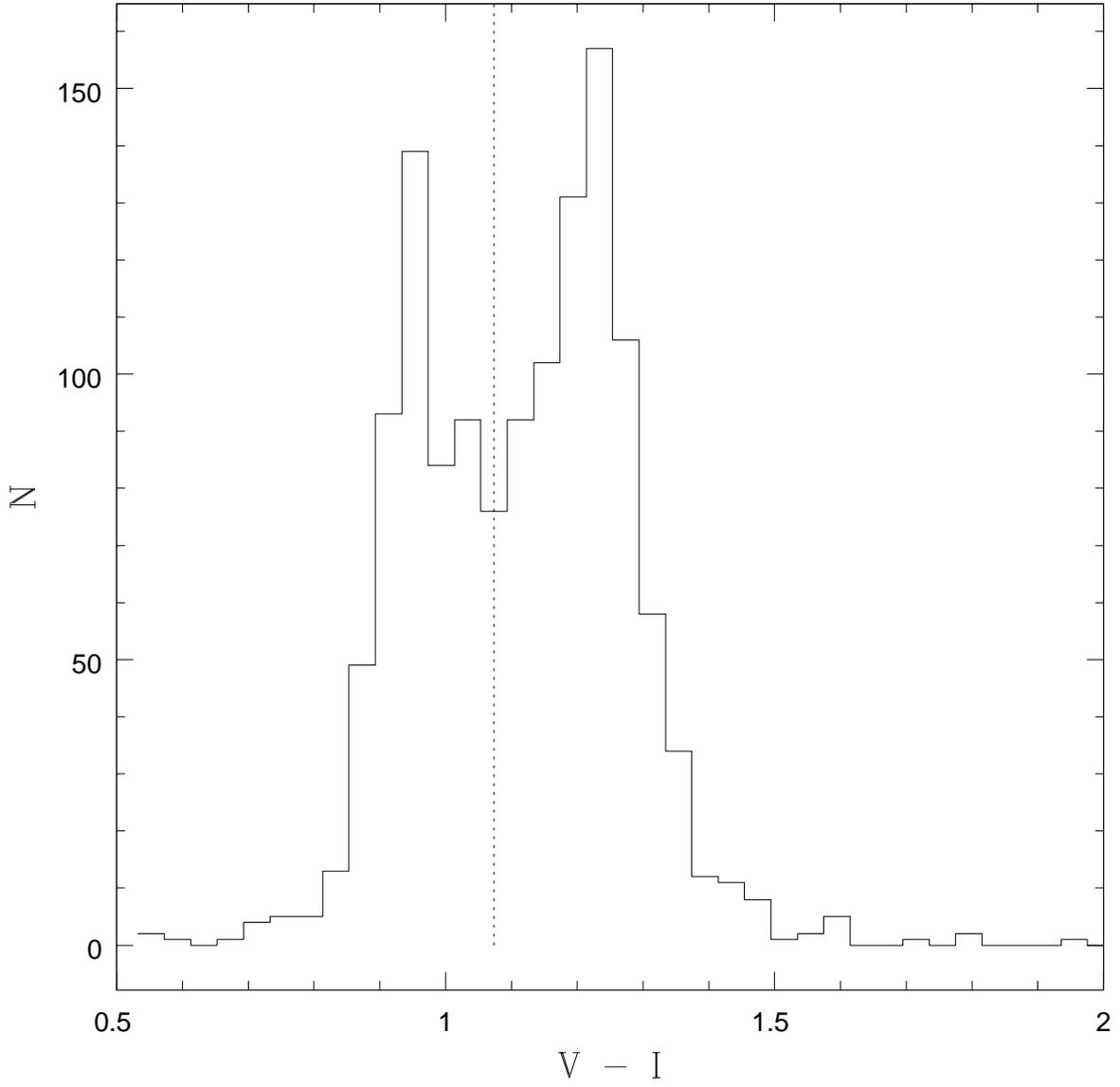}
\caption{Colour distribution of observed globular cluster candidates in M87. The vertical dotted line marks the separation between red and blue globular clusters.}
  \label{fig:col_dis}
\end{figure}

Details regarding the identification of globular cluster candidates and the measurement of cluster sizes via \citet{king62} model fitting can be found in WSH12. The final candidate list contained 1290 globular clusters identified with high confidence within projected distances $R_{gc} < 10$ kpc. To separate the population into red clusters and blue clusters, we consider the colour distribution of the globular cluster candidates in Figure \ref{fig:col_dis}. The vertical dotted line in Figure \ref{fig:col_dis} at (V-I) = 1.07 marks a local minimum in the colour distribution separating the blue globular cluster population (left of vertical line) and the red globular cluster population (right of vertical line).

\begin{figure}[tbp]
\centering
\includegraphics[width=\columnwidth]{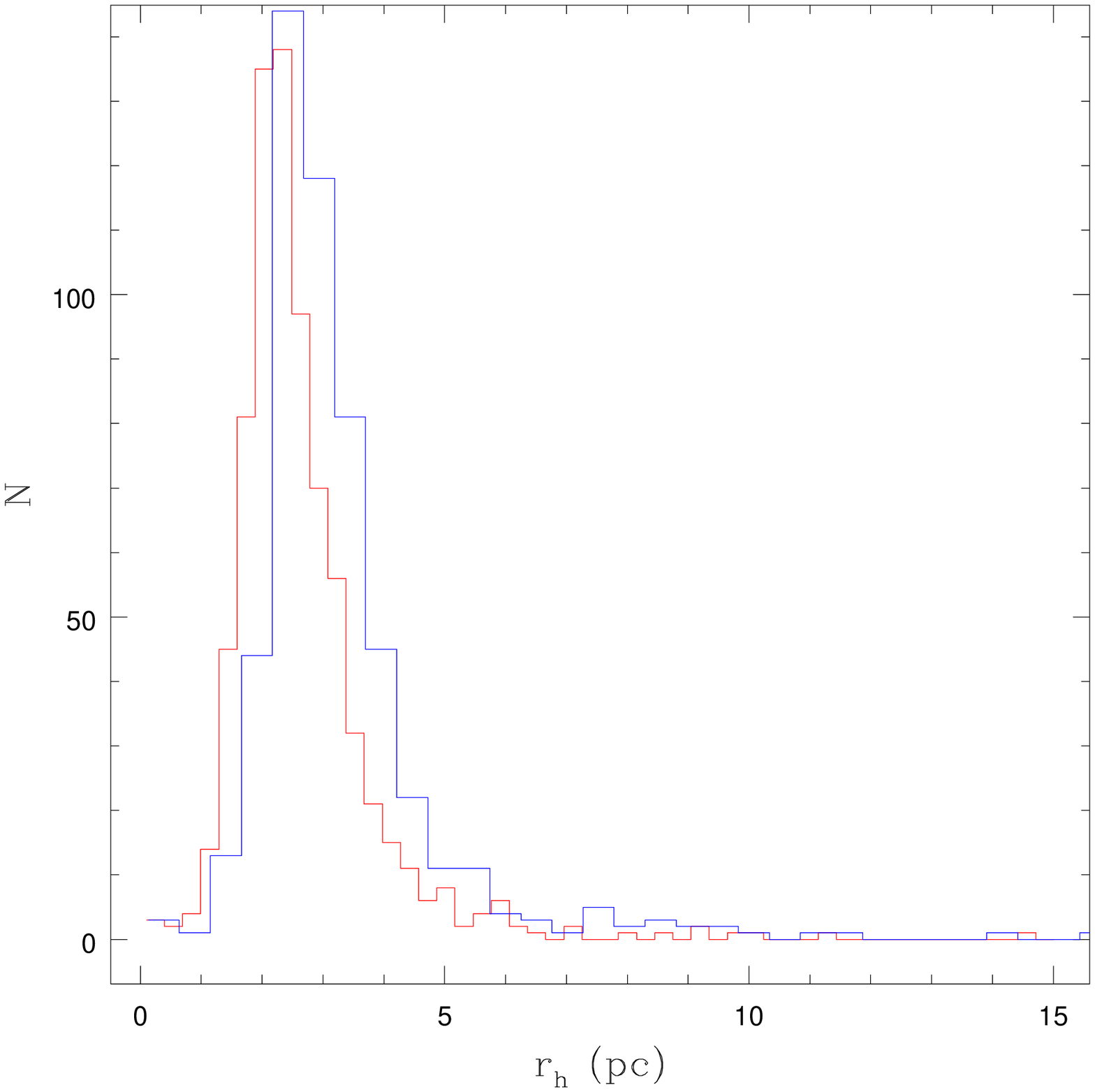}
\caption{Effective radius ($r_h$) distribution of observed globular cluster candidates in M87. The red histogram corresponds to red globular clusters and the blue histogram corresponds to blue globular clusters.}
  \label{fig:rh_hist_obs}
\end{figure}

In Figure \ref{fig:rh_hist_obs} we compare red and blue cluster sizes by showing the distribution of the average effective radius of each cluster in the V and I bands, as defined by standard \citet{king62} profiles. In general, red globular clusters are smaller than blue globular clusters. We find the overall mean red cluster size (2.65 pc) to be approximately $30 \%$ (0.71 pc) smaller than the mean blue cluster size (3.36 pc). The difference of $30 \%$ is slightly higher than both the difference found by \citet{madrid09} for M87 ($24 \%$) and the mean difference of $20 \%$ found in the literature for all galaxy types. This difference is a result of our strict selection criteria when identifying globular cluster candidates (see WSH12), as our cluster sizes are comparable to those of \citet{peng09} and \citet{madrid09}. However, since the simulation outlined in Section 3 is tailored to fit our observational dataset, the discrepancy will not affect our subsequent analysis. A Kolmogorov-Smirnov (KS) test indicates that the $r_h$ distribution of the red and blue clusters have essentially zero ($2 \times 10^{-23}$) probability of being drawn from the same distribution.

\begin{figure}[tbp]
\centering
\includegraphics[width=\columnwidth]{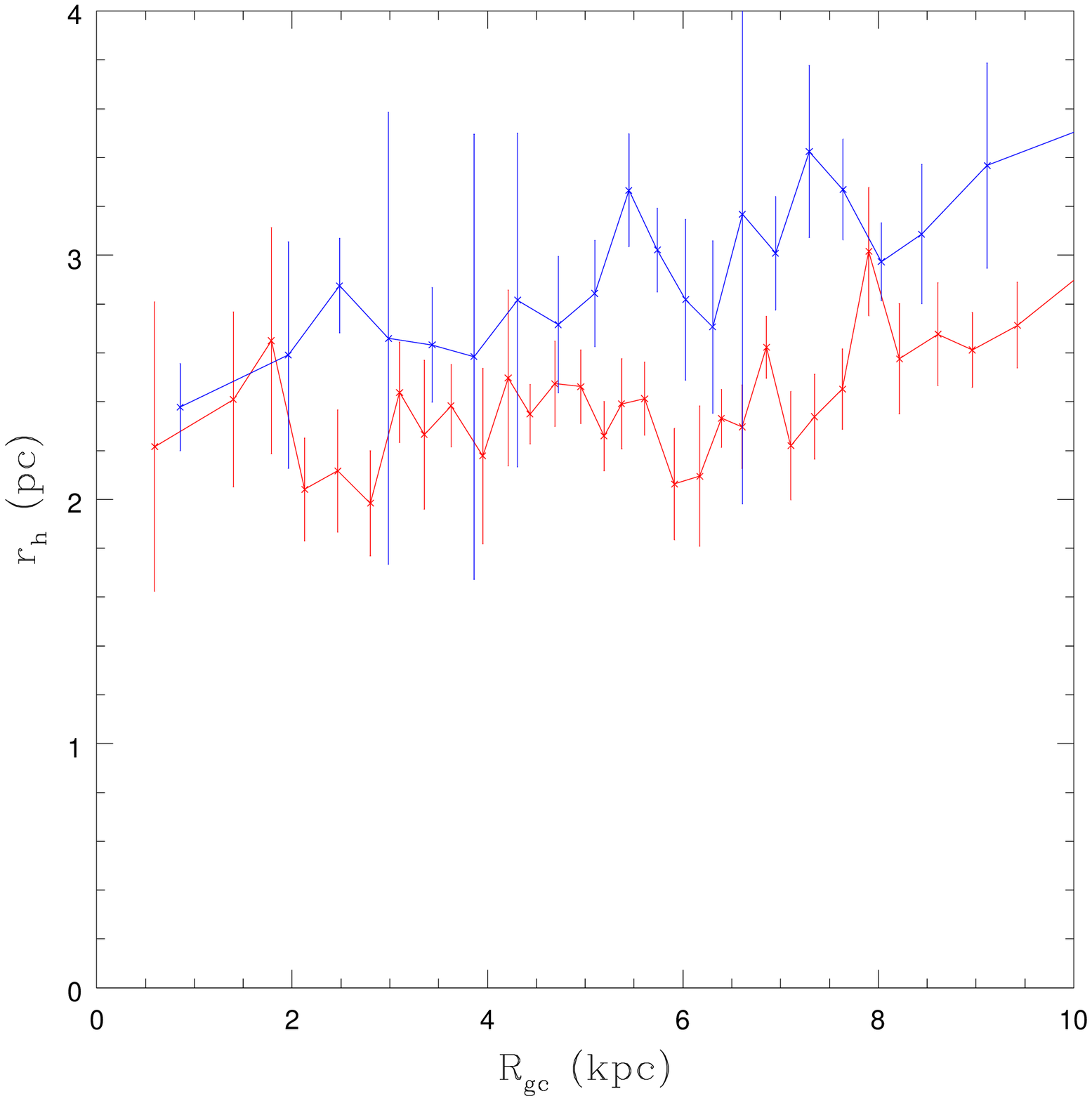}
\caption{Relationship between median effective radius ($r_h$) and projected galactocentric distance ($R_{gc}$) for observed globular cluster candidates in M87. The red line corresponds to red globular clusters and the blue line corresponds to blue globular clusters. Each bin contains 25 globular cluster candidates. Error bars represent the standard error $\sigma/\sqrt(n)$ as given by \citet{harris10}.}
  \label{fig:rh_rgc_obs}
\end{figure}

A plot of median effective radius versus projected galactocentric distance is shown in Figure \ref{fig:rh_rgc_obs}. Here we see that red clusters appear to be smaller than blue clusters at all radii. 

\section{Simulation \label{sthree}}

Our simulation of a globular cluster population within M87 (outlined in WSH12) allows for the simulated clusters to have the same characteristics as the observed globular cluster population. A position ($R$, $\theta$, $\phi$), velocity ($v_r$, $v_\theta$, $v_\phi$), mass (M), and central concentration (the King-model c-value \citep{king66}) was assigned to each cluster. Cluster mass was taken from the near-universal globular cluster luminosity function \citep[e.g.][] {brodie06}, with luminosities converted to mass by assuming a mass-to-light ratio of 2 (e.g. \citet{mclaughlin05}). The simulation allowed for clusters to have a range of central concentrations, with the distribution set equal to that of the Milky Way cluster population (Harris 1996 (2010 Edition)).

The radial velocity dispersion of the globular clusters was set equal to the line of site velocity dispersion of the observed M87 clusters \citep{cote01}. In WSH12, in order for the simulation to match the observed relationship between cluster size and projected galactocentric distance we allowed for outer region clusters to have increased orbital anisotropy such that they prefer radial orbits. This was included in the simulation presented here, and was used to calculate the tangential velocity dispersion of the simulated clusters.

Our simulation also allowed for red and blue clusters to have two separate spatial distributions. These distributions were taken from \citet{harris09}, who found that the projected radial profile of the blue and red globular cluster subpopulations could be fit with a standard Hubble profile relating density ($\sigma_{cl}$) to projected distance ($R$):

\begin{equation}\label{hprof}
\sigma_{cl}(R)=\sigma_0 / (1+\frac{R}{R_0})^{-a}
\end{equation}

The appropriate values for $R_o$ and $a$ are $2.0'$  and 1.8 for blue clusters and $1.2'$ and 2.1 for red clusters.  A three dimensional radial profile was then obtained by assuming that the globular cluster population of M87 is spherically symmetric, and integrating Equation \ref{hprof} over a spherical volume. This results in our simulated projected radial profile matching Equation \ref{hprof} within uncertainty. The angular distribution was also assumed to be spherically symmetric. 

Exactly 6000 globular clusters were simulated so the number of clusters within 10 kpc of M87 would be the same as the observed dataset. Clusters were simulated out to 100 kpc following the observed profile to ensure that larger distant clusters could be projected to small projected distances. $40 \%$ of the clusters were designated as ``red" clusters and $60 \%$ were designated as ``blue" clusters such that the ratio of blue clusters to red clusters is in agreement with the data in \citet{harris09}.

Using the simulated parameters of each cluster, we calculate the theoretical tidal radius of each cluster as derived by \cite{bertin08}:

\begin{equation} \label{rt}
r_t=(\frac{GM}{\Omega^2\upsilon})^{1/3}
\end{equation}

\noindent Where $\Omega$, $\kappa$ and $\upsilon$ are defined as:

\begin{equation}
\Omega^2=(d\Phi_G(R)/dR)_{R_p}/R_p
\end{equation}
\begin{equation}
\kappa^2=3\Omega^2+(d^2\Phi_G(R)/dR^2)_{R_p}
\end{equation}
\begin{equation}
\upsilon=4-\kappa^2/\Omega^2
\end{equation}

\noindent $\Phi_G$ is the galactic potential determined from the mass profile of M87, $R_p$ is the cluster's perigalactic distance, $\Omega$ is the orbital frequency of the cluster and $\kappa$ is the epicyclic frequency of the cluster at $R_p$. Tidal truncation is the only effect on cluster size considered in this simulation, which assumes that all clusters are tidally filling. It should be noted that whether a cluster is red or blue has no effect on the calculation of its tidal radius. The mass profile of M87 was taken from \citet{mclaughlin99}. To best compare with our observations, tidal radii are converted to effective radii by assuming each simulated cluster can be represented by a \citet{king62} model. 

\begin{figure}[tbp]
\centering
\includegraphics[width=\columnwidth]{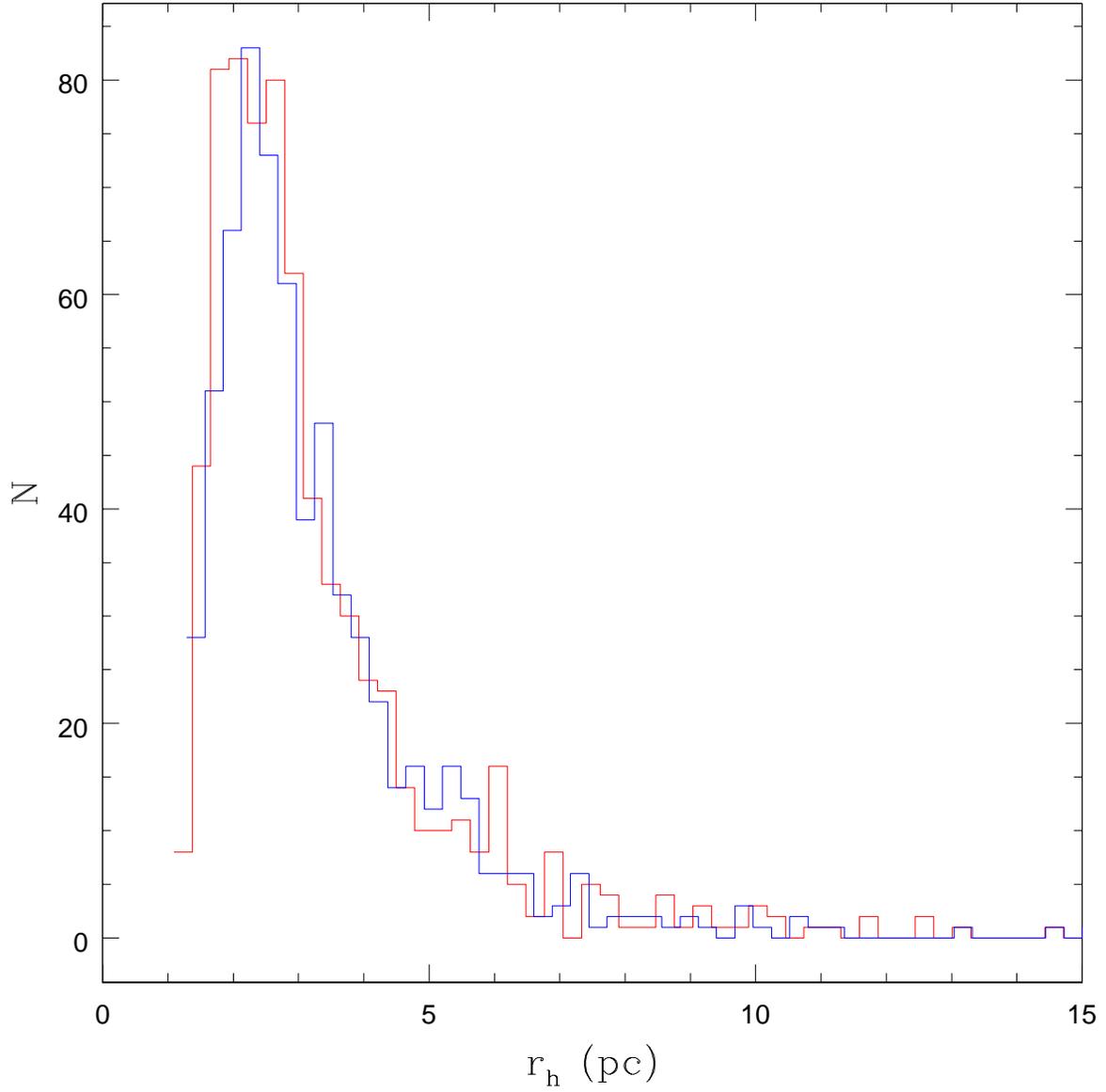}
\caption{Effective radius ($r_h$) distribution of simulated globular clusters in M87. The red histogram corresponds to red globular clusters and the blue histogram corresponds to blue globular clusters.}
  \label{fig:rh_hist_mod}
\end{figure}

To determine if the size difference between red and blue globular clusters is due to projection effects we re-plot Figure \ref{fig:rh_hist_obs}, but this time with the simulated cluster population in Figure \ref{fig:rh_hist_mod}. It should be noted that Figure \ref{fig:rh_hist_mod} only contains simulated clusters with projected distances within 10 kpc, just as the observations do. Here we see that the mean simulated red globular cluster size (3.31 pc) and the mean blue globular cluster size (3.36 pc) are essentially the same. Scatter about the mean value is simply due to the distribution of concentrations, velocities, and masses allowed by our model. The KS test cannot reject the hypothesis of the two samples being drawn from the same parent distribution with more than a $79 \%$ confidence level, a much stronger agreement than the observed distributions.

While it remains true that outer region clusters experience weaker tidal truncation than inner region clusters, the lack of a size difference in our simulation indicates that the projection of larger, outer region blue clusters is not a strong enough effect to explain the observed size difference between the mean red and blue cluster sizes. The fact that outer region clusters have preferentially radial orbits will also contribute to the lack of a size difference in our simulation as some outer region clusters will still travel deep into the tidal field of M87, increasing the effect of tidal truncation.

\begin{figure}[tbp]
\centering
\includegraphics[width=\columnwidth]{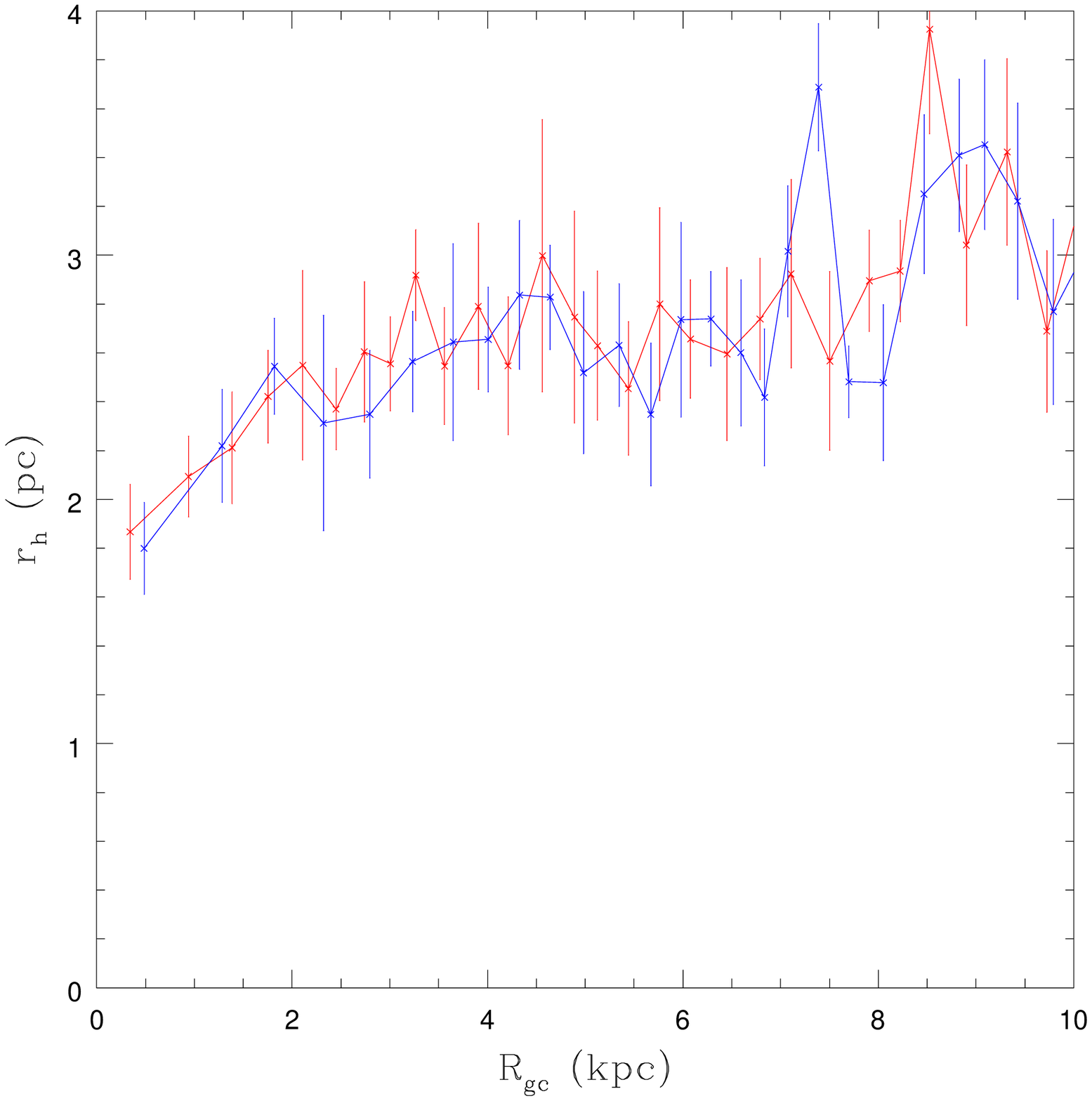}
\caption{Relationship between effective radius ($r_h$) and projected galactocentric distance ($R_{gc}$) for simulated globular clusters in M87. The red line corresponds to red globular clusters and the blue line corresponds to blue globular clusters. Each bin contains 25 simulated globular clusters. Error bars represent the standard error $\sigma/\sqrt(n)$ as given by \citet{harris10}.}
  \label{fig:rh_rgc_mod}
\end{figure}

We further explore the influence of projection effects by plotting the relationship between cluster size and projected galactocentric distance for red and blue clusters in Figure \ref{fig:rh_rgc_mod}. At a given projected galactocentric distance, again the median red cluster size is equal to the median blue cluster size. The results of our simulation are in direct disagreement with the observations in Figures \ref{fig:rh_hist_obs} and \ref{fig:rh_rgc_obs}, which showed that at all projected distances blue clusters are larger than red clusters. This statement is still true even if the model cluster population had an isotropic distribution of orbits. Therefore, the size difference between red and blue globular clusters cannot be explained by just the projection of two sub-populations that only differ by their spatial distributions.

\section{Summary \label{sfour}}

While projection effects have long been one possible explanation as to why red clusters are smaller than blue clusters, we show that the size difference between red and blue clusters cannot be explained by projection effects alone, and that another mechanism is playing a more dominant role. In fact, this work suggests that red clusters must either under-fill their tidal radius or have different orbital parameters than the blue population in order to be sufficiently smaller than blue clusters. Specifically, if red clusters had preferentially radial orbits compared to blue clusters, it would result in red clusters having smaller tidal radii than blue clusters at all galactocentric distances. We suggest that the metallicity of a globular cluster must play an important role in determining exactly how large or small a globular cluster can be, through either initial conditions or later dynamical and stellar evolution \citep{jordan04, jordan05, harris09b, downing12}. 

Applying our model to other galaxies with different galactic potentials and globular cluster systems will confirm whether this conclusion applies to all galaxies, all early-type galaxies, or just M87. Future work will explore the assumptions made in our simulation, including the relationship between tidal radii and effective radii and additional differences between the red and blue clusters in our model, such as the possibility of the sub-populations having different orbital anisotropy profiles. It is also worth exploring what influence globular clusters that do not fill their tidal radius have on our comparison, as some studies find cluster sizes increase at a much shallower rate at large galactocentric distances (or not at all) compared to our model \citep[e.g.][] {gomez07, spitler06, harris09b, harris10}. Finally, simulations of clusters with a range of metallicities, similar to those of \citet{sippel12} which find the size difference can be attributed to the effect that metallicity has on stellar evolution and dynamical effects, will also shed a significant amount of light on this issue.

\section{Acknowledgements}

JW would like to acknowledge funding through the A. Boyd McLay Ontario Graduate Scholarship. WEH and AS acknowledge financial support through research grants from the Natural Sciences and Engineering Research Council of Canada. 



\end{document}